\documentstyle[twoside,fleqn,espcrc2,psfig]{article}
\newcommand{\beq}{\begin{equation}}
\newcommand{\eeq}{\end{equation}}
\newcommand{\bed}{\begin{displaymath}}
\newcommand{\eed}{\end{displaymath}}
\newcommand{\bea}{\begin{eqnarray}}
\newcommand{\eea}{\end{eqnarray}}

\renewcommand{\b}{\beta}

\newcommand{\ra}{\rightarrow}

\newcommand{\AmS}{{\protect\the\textfont2
  A\kern-.1667em\lower.5ex\hbox{M}\kern-.125emS}}

\title{Center Vortices and the Asymptotic String Tension}

\author{L. Del Debbio\address{CPT, CNRS Luminy, Case 907, F-13288 Marseille
Cedex 9, France}, M. Faber\address{Inst. f{\"u}r Kernphysik, Tech. Univ.
Wien, A-1040 Vienna, Austria}, J. Greensite\address{The Niels Bohr Institute,
DK-2100 Copenhagen \O, Denmark}
\thanks{Talk presented by J. Greensite.  Work supported by 
Carlsbergfondet, and by the U.S. Department 
of Energy under Grant No. DE-FG03-92ER40711.},
\v{S}. Olejn{\'\i}k\address{Inst. of Phys., Slovak Acad. of Sci., 
SK-842 28 Bratislava, Slovakia} }

\begin{document}

\begin{abstract}
   We present a method for locating center vortices (``fluxons'') in 
thermalized lattice gauge field configurations.  We find evidence, in
lattice Monte Carlo simulations, that the asymptotic string tension of
fundamental-representation Wilson loops is due to fluctuations in the
number of center vortices linking those loops.
\end{abstract}

\maketitle

\section{Introduction}

   In this contribution I would like to discuss a variety of numerical
data, obtained recently by our group, which supports the Center Vortex 
Theory of confinement.  This theory was proposed, in various
forms, by 't Hooft \cite{tHooft2}, Mack \cite{Mack} and by 
Nielsen and Olesen \cite{CopVac} (the ``Copenhagen Vacuum''), in the
late 1970's.  Space limitations do not allow me to actually display very much 
of the relevant data here; for this I must refer the interested reader 
to some other recent conference proceedings \cite{Zako}.

   The most popular theory of quark confinement is the abelian projection
theory proposed by 't Hooft \cite{tHooft1}.  In past years our group has 
been highly critical of this theory (as well as the center vortex theory), 
on the grounds that it fails to explain the existence of a linear potential 
between higher representation
quarks in the Casimir scaling regime \cite{lat96,Cas1}.    
This failure is very significant, because it is in the Casimir regime that
the confining force replaces Coulombic behavior, and in fact it is only
in this regime that the QCD string has been well studied numerically.
If we don't understand Casimir scaling, then we don't really understand how
flux tubes form. 

   A possible response to this criticism is simply to admit 
that the formation of flux tubes, at intermediate distances, remains to be 
understood, but that the abelian projection theory is nonetheless valid at 
very large distance scales, where Casimir scaling breaks down and
color screening sets in.  I will argue that 
there may be some truth to this response, but that the confining configurations
relevant to this asymptotic regime seem to be $Z_N$ vortices, rather than 
abelian monopoles.     
   
\section{Center Dominance}

   I will begin with the phenomenon of ``center dominance,''
which we reported, at the Lattice 96 meeting last year \cite{lat96}, as 
part of a critique of the abelian projection theory.  
The idea is as follows:  In an $SU(2)$ lattice gauge theory, begin 
by fixing to maximal abelian gauge \cite{Kronfeld}. Then go one step further,
using the remnant $U(1)$ symmetry to bring the abelian link variables
$A = \mbox{diag}[ e^{i\theta}, e^{-i\theta}]$
as close as possible to the $SU(2)$ center elements $\pm I$, by
maximizing $<\cos^2 \theta>$, leaving a remnant $Z_2$ symmetry.
This is the {\bf (indirect) Maximal Center Gauge} (the center
is maximized in $A$, rather than directly in the full link variables
$U$).  We then define, at each link,
$Z \equiv \mbox{sign}(\cos\theta) = \pm 1$
which is easily seen to transform like a $Z_2$ gauge field under the
remnant $Z_2$ symmetry.  ``Center Projection'' is the replacement 
$U \rightarrow Z$ of the full link variables by the center variables; we can
then calculate Wilson loops, Creutz ratios, etc. with the center-projected
$Z$-link variables.  What we found was the following \cite{lat96}:
\begin{itemize}
\item[1.] Center-projected Creutz ratios $\chi(R,R)$ scale very nicely
with $\b$, following the usual prediction of asymptotic freedom.  Moreover, 
at fixed $\beta$, they are nearly $R$-independent, indicating that the 
Coulombic contribution is suppressed, and only the
constant confining force remains.
\item[2.] If the $Z$ variable is factored out of the abelian links, 
and loops are computed from the $A/Z$ variables, the string tension
disappears.
\end{itemize}
The fact that the $Z$ variables seem to carry most of the information
about the confining force is what we mean by ``center dominance.''

   The only excitations of $Z_2$ lattice gauge theory with non-zero action
are ``thin'' $Z_2$ vortices, which have the topology of a surface
(one lattice spacing thick) in D=4 dimensions.  We will call the
$Z_2$ vortices, of the center projected $Z$-link configurations, 
``Projection-vortices'' or just {\bf P-vortices}. 
These are to be distinguished from the hypothetical ``thick'' center 
vortices, which might exist in the full, unprojected $U$ configurations.  
The first question to ask is whether the presence or absence of P-vortices, in
a given center-projected lattice, is correlated with the  
confining properties of the corresponding unprojected lattice.

\section{Vortex-Limited Wilson Loops}

   We will say that a plaquette is ``pierced'' by a P-vortex if, upon
going to maximal center gauge and center-projecting, the projected
plaquette has the value $-1$.  Likewise, a given lattice surface is 
pierced by $n$ P-vortices if $n$ plaquettes of the surface are pierced
by P-vortices.

   In a Monte Carlo simulation, the number of P-vortices piercing
the minimal area of a given loop $C$ will, of course, fluctuate.  Let
us define $W_n(C)$ to be the Wilson loop evaluated on a sub-ensemble
of configurations, selected such that precisely $n$ P-vortices, in
the corresponding center-projected configurations, pierce the minimal area
of the loop.  It should be emphasized here that the center projection
is used only to select the data set.  The Wilson loops themselves are
evaluated using the full, {\it unprojected} link variables.  
Then, if the presence or absence of P-vortices in the 
projected configuration is unrelated to the confining properties of
the corresponding unprojected configuration, we would expect
\beq
\chi_0(I,J) \approx \chi(I,J)
\eeq
at least for large loops.

   The result of this test is shown in Fig. 1.  Quite contrary to
our original expectations, the confining force vanishes if P-vortices are
excluded.  This does not necessarily mean that the confining
configurations of $SU(2)$ lattice gauge theory are thick center
vortices.  It does imply, however, that the presence or absence
of P-vortices in the projected gauge field is strongly correlated with
\begin{figure}
\centerline{\hbox{\psfig{figure=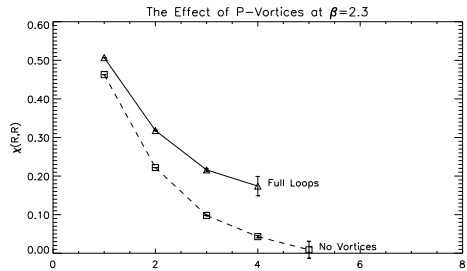,width=6.5cm}}}
\caption[chi]{Creutz ratios $\chi_0(R,R)$ extracted from loops with
no P-vortices, as compared to the usual Creutz ratios $\chi(R,R)$,
at $\beta=2.3$.}
\end{figure}   
the presence or absence of confining configurations (whatever they
may be) in the unprojected gauge field. 

   The next question is whether these confining configurations are,
in fact, thick center vortices.  If they are, then a short argument
(see ref. \cite{Zako,Us}) leads to the prediction that 
\beq
W_n[C]/W_0[C] \ra (-1)^n
\eeq
 as the loop size increases.
Figure 2 shows the ratio $W_1/W_0$, which seems to confirm this
prediction; we have other data showing that $W_2/W_0 \ra +1$.
\begin{figure}
\centerline{\hbox{\psfig{figure=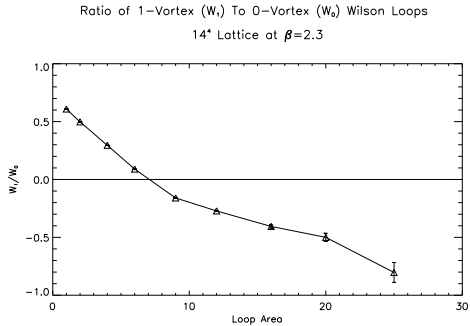,width=6.5cm}}}
\caption[vtex1]{Ratio of the 1-Vortex to the 0-Vortex Wilson loops,
$W_1(C)/W_0(C)$, vs. loop area at $\beta=2.3.$}
\end{figure}
We have also considered loops pierced by only even, or only odd, numbers of 
P-vortices, and found that the string tension in those cases vanishes,
and that $W_{odd}[C]/W_{even}[C]\ra -1$.  This data, in conjunction
with center dominance, is a strong
indication that P-vortices are correlated with thick center vortices,
and that these thick vortices are responsible for the confining force.
It is also consistent, we believe, with related results reported by 
Kov\'{a}cs and Tomboulis at this meeting \cite{TK}.

\section{Latest Results}

  I would like to briefly mention some further developments \cite{Zako}:
\begin{itemize}
\item[1.] We have introduced a {\bf (direct) Maximal Center Gauge},
which brings the entire link variable (not just the abelian part)
as close as possible to $\pm I$.  In this gauge there is not only scaling,
but also very close numerical agreement of
the center-projected string tension with currently accepted values
for the asymptotic string tension.
\item[2.] We have found, in the (indirect) maximal center gauge, that almost
all monopoles found in the abelian projection lie on P-vortices,
and that virtually all of the excess field
strength of (unprojected) monopole cubes, above the lattice average, is 
directed along the P-vortices.  Monopoles would appear to be an artifact
of the abelian projection; they are condensed because the underlying vortices
from which they emerge are condensed.
\end{itemize}

  Finally, there is the issue of the Casimir scaling of 
higher-representation string tensions, in the intermediate distance regime.  
We have not forgotten the point that Casimir scaling, whose importance we 
have often emphasized \cite{lat96,Cas1}, does not seem to be explained by 
the center vortex theory  \cite{Us}. Nor has this point gone unnoticed 
by other people at this meeting \cite{Pierre}.  Very recently we 
have found a possible explanation for Casimir scaling within the 
framework of the vortex theory.  This explanation will be reported 
elsewhere.

\end{document}